# Identification Of Outliers In Oxazolines And Oxazoles High Dimension Molecular Descriptor Dataset Using Principal Component Outlier Detection Algorithm And Comparative Numerical Study Of Other Robust Estimators

Doreswamy[1] and Chanabasayya .M. Vastrad[2]

[1]Department of Computer Science, Mangalore University,
Mangalagangotri-574 199, Karnataka, India
[1]Doreswamyh@yahoo.com
[2]channu.vastrad@gmail.com

## ABSTRACT

*From the past decade outlier detection has been in use. Detection of outliers is an emerging topic and is having robust applications in medical sciences and pharmaceutical sciences. Outlier detection is used to detect anomalous behaviour of data. Typical problems in Bioinformatics can be addressed by outlier detection. A computationally fast method for detecting outliers is shown, that is particularly effective in high dimensions. PrCmpOut algorithm make use of simple properties of principal components to detect outliers in the transformed space, leading to significant computational advantages for high dimensional data. This procedure requires considerably less computational time than existing methods for outlier detection. The properties of this estimator (Outlier error rate (FN), Non-Outlier error rate(FP) and computational costs) are analyzed and compared with those of other robust estimators described in the literature through simulation studies. Numerical evidence based Oxazolines and Oxazoles molecular descriptor dataset shows that the proposed method performs well in a variety of situations of practical interest. It is thus a valuable companion to the existing outlier detection methods.*

## KEYWORDS

*Robust distance, Mahalanobis distance, median absolute deviation, Principal Components Analysis, Projection-Persuit ,*

## 1. INTRODUCTION

Accurate detection of outliers plays an important role in statistical analysis. If classical statistical models are randomly applied to data containing outliers, the results can be deceptive at best. In addition , An outlier is an observation that lies an abnormal distance from other values in a data from a dataset and their identification is the main purpose of the investigation. Classical methods based on the mean and covariance matrix are not often able to detect all the multivariate outliers in a descriptor dataset due to the masking effect [1] , with the result that methods based on classical measures are inappropriate for general use unless it is certain that outliers are not present. Erroneous data are usually found in several situations, and so robust methods that detect or down weight outliers are important methods for statisticians. The objective of this examination is to provide an detection of outliers, prior to whatever modelling process is visualized. Sometimes detection of outliers is the primary purpose of the analysis, other times the outliers need to be removed or down weighted prior to fitting non robust models. We do not distinguish between the various reasons for outlier detection, we simply aim to inform the analyst of observations that are considerably different from the majority. Our procedures are therefore exploratory, and applicable to a wide variety of settings.

Most methods with a high resistance to outliers are computationally exhaustive; not accordingly, the availability of cheap computing resources has enabled this field to develop significantly in recent years.

Among other proposals, there currently exist a wide variety of statistical models ranging from regression to principal components [2] that can include outliers without being excessively influenced, as well as several algorithms that explicitly focus on outlier detection.

There are several applications where multi-dimensional outlier identification and/or robust estimation are important. The field of Computational drug discovery, for instance, has recently received a lot of concentration from statisticians (e.g. the project Bioconductor, http://www.bioconductor.org). Improvements in computing power have allowed pharmacists to record and store extremely large databases of information. Such information likely to contain a fair amount of large errors, however, so robust methods is needed to prevent these errors from influencing the statistical model. Undoubtedly, algorithms that take a long time to compute are not perfect or even practical for such large data sets.

Also, there is a further difficulty encountered in molecular descriptor data. The number of dimensions is typically several orders of importance larger than the number of observations, leading to a singular covariance matrix, so the majority of statistical methods cannot be applied in the usual way. As will be discussed later, this situation can be solved through singular value decomposition but it does require special attention. It can thus be seen that there are a number of important applications in which current robust statistical models are impractical.

## 2. MATERIALS AND ALGORITHAMS

### 2.1 The Data Set

The molecular descriptors of 100 Oxazolines and Oxazoles derivatives [30-31] based H37Rv inhibitors analyzed. These molecular descriptors are generated using Padel-Descriptor tool [32]. The dataset covers a diverse set of molecular descriptors with a wide range of inhibitory activities against H37Rv.

### 2.2 A Brief Overview of Outlier Detection

There are two basic ways to outlier detection – distance-based methods, and projection pursuit. Distance-based methods aim to detect outliers by computing a measure of how far a particular point is from the centre of the data. The common measure of "outlyingness" for a data point $x_i \in \mathbb{R}^p$ , $i = 1, \ldots, n$, is a robust version of the Mahalanobis distance,

$$RD_i = \sqrt{(x_i - T)^T C^{-1}(x_i - T)} \qquad (1)$$

Where $T$ is a robust measure of location of the descriptor data set $X$ and $C$ is a robust estimate of the covariance matrix. Problems faced by distance-based methods include (i) obtaining a reliable estimate of **C** in addition to (ii) how large $RD_i$ should be before a point is classified as outlying. This focuses the devoted connection between outlier detection and robust estimation – the latter is required as part of the prior. Retrieving good robust estimators of $T$ and $C$ are essential for distance-based outlier detection methods. It is then crucial to find a metric (based on $T$ and $C$) set apart outliers from regular points. The final separation boundary commonly depends on user-specified penalties for misclassification of outliers as well as regular points(inliers).

### 2.2.1 Robust Estimation as Main Goal

A simple robust estimate of location is the coordinatewise median. This estimator is not orthogonally equivariant (does not transform correctly under orthogonal transformations), but if this property is important, the L1 median should be used alternatively, expressed as

$$\hat{\mu} = \underset{\mu \in R^p}{argmin} \sum_{i=1}^{n} \|x_i - \mu\|, \qquad (2)$$

where $\|\cdot\|$ stands for the Euclidean norm. The L1 median has maximal breakdown point, and a fast algorithm for its computation is given in [3] .

A simple robust estimate of scale is the MAD (median absolute deviation), expressed for a dataset $\{x_1,....,x_n\} \subset \mathbb{R}$ as

$$MAD(x_1, ...., x_n) = 1.4826 \cdot \underset{j}{med} \left| x_j - \underset{i}{med} \, x_i \right|. \quad (3)$$

More complex estimators of location and scale are presented by the class of S-estimators [4] , expressed as the vector $T$ and positive definite symmetric matrix $C$ that satisfy

$$\min|C| \qquad s.t. \qquad \frac{1}{n} \sum_{i=1}^{n} \rho(d_i/c) = b_0,$$

$$d_i = \sqrt{(x_i - T)^T C^{-1}(x_i - T)} \, , \qquad (4)$$

where $\rho(\cdot)$ is a non-decreasing function on $[0,\infty)$ , and **c** and $b_0$ are tuning constants that can be as one chosen to support particular breakdown properties. It is commonly easier to work with $\psi = \partial \rho/\partial d$ after all $\psi$ has a root where $\rho$ has minimum.

Distance based algorithms that follow robust estimation as a main goal – without explicit outlier detection – contain the OGK estimate [5] the minimum volume ellipsoid (MVE) and minimum covariance determinant (MCD)[6-7]

MCD tries to find the covariance matrix of minimum determinant containing at least h data points, where h determines the robustness of the estimator; it should be at least $(n + p - 1)/2$. The MCD and MVE are instances of S-estimators with non-differentiable $\rho(d_i)$ because $\rho(d_i)$ is either 0 or 1. MCD shows good performance on data sets with low dimension but on larger data sets the computational stress can be restrictive – the accurate solution requires a combinational search. In the latter case good starting points need to be acquired, accommodating an approximately correct method. Equivariant procedures of acquiring  these starting points, however, are based on subsampling methods and the number of subsamples  needed to acquire an tolerable level of accuracy increases swiftly with dimension. Rousseeuw and van Driessen (1999) developed a faster version of MCD which was a considerable advancement, but is still quite computationally exhaustive. The OGK estimator [5] is located on pair wise robust estimates of the covariance. Gnanadesikan and Kettenring (1972) computed a robust covariance estimate for two variables $X$ and $Y$ based on the uniqueness

$$COV(X,Y) = \frac{1}{4}(\sigma(X+Y)^2 - \sigma(X-Y)^2), \qquad (5)$$

where σ is a robust estimate of the variance. The matrix build from these pair wise estimates will not necessarily be positive semi definite, so Maronna and Zamar (2002) carried out by performing an eigen decomposition of this matrix. After all the variables in eigenvector space are orthogonal, the covariances are zero and it is enough  to get robust variance estimates of the data projected onto each eigenvector direction. The eigenvalues are then restored with these robust variances, and the eigenvector conversion is applied in reverse to give in a positive semidefinite robust covariance matrix. If the original data matrix is robustly scaled (every component divided by its robust variance), the OGK will be scale invariant. This method can be iterated, despite Maronna and Zamar (2002) find this is not consistently better. Maronna and Zamar (2002) in addition to find that using weighted estimates is to some extent better, in which case the observations are weighted according to their robust distances $d$ as scaled by the robust covariance matrix. They concern a weighting function of the form $I(d < d_0)$ where $I(\cdot)$  is the indicator function and $d_0$ is taken to be

$$d_0 = \frac{\chi_p^2(\beta) \, med(d_1, \ldots, d_2)}{\chi_p^2(0.5)}, \qquad (6)$$

where  $\chi_p^2(\beta)$ is the $\beta$ -quantile of the $\chi_p^2$ distribution. Observations thus admit full weight except that their robust distance d > $d_0$, in which case they admit zero weight. Maronna and Zamar (2002) note that the robust distances $d$ can be swiftly calculated in the eigenvector space without the requirement for matrix inversion because the $p$ components are orthogonal in this space. That is,

$$d_i = \sum_{j=1}^{p} \left( \frac{z_{ij} - \mu(Z_j)}{\sigma(Z_j)} \right)^2, \quad i = 1, \ldots, n, \qquad (7)$$

where $z_{ij}$ are the data in the space of eigenvectors , $Z_j$ are the $p$  components in this space, $\mu$ is robust location estimate and $\sigma$ is a robust variance estimate.

### 2.2.2 Explicit Outlier Detection

Continuing robust estimation to outlier detection needs some knowledge of the distribution of robust distances If $\mathbf{X}$ pursues a multivariate normal distribution, the squared classic Mahalanobis distance (based upon the sample mean and covariance matrix) pursues a $\chi_p^2$ distribution [8]. In addition to, if robust estimators $\mathbf{T}$ and $\mathbf{C}$ are applied to a large data set in which the non-outliers are normally distributed. Hardin and Rocke (2005) discovered that the squared distances could be expressed by a scaled $\mathbf{F}$ - distribution. After all, for non-normal data, it is not clear how the outlier boundary should be driven to give optimal classification rates. These regards form the basis for the use of $d_0$ by Maronna and Zamar (2002). The complete change of equation (6) helps the distribution of $d_i$'s be like that of $\chi_p^2$ for non-normal original data, heading to better results for the cutoff value than simply $\chi_p^2(\beta)$.

Hopeful algorithms that focus on detection of outliers contain rocke-Estimator [25], sfast-Estimator[26],M-Estimator[27],MVE-Estimator[28],NNC-Estimator [29] ,BACON[9], PCDist[10-11] ,sign1[12-13] and sign2[14], Outlier identification using robust (mahalanobis) distances based on robust multivariate location and covariance matrix [15]. BACON and robust multivariate location and covariance matrix are distance-based and in an appropriate, direct the larger part of computational effort toward obtaining robust estimators $\mathbf{T}$ and $\mathbf{C}$ . BACON begins with a small subset of observations believed to be outlier-free, to which it iteratively adds points that have a small Mahalanobis distance based on T and $\mathbf{C}$ of the current subset. One reason that builds MCD unreliable for high $\mathbf{p}$ is that its contamination bias evolves very swiftly with p [16]. Robust multivariate location and covariance matrix aims to reduce the computational burden by subdividing the data into cells and running MCD on each cell, i.e. reducing the number of observations that MCD performs on, with the same number of dimensions. It then associates the results from each cell to yield a beginning point for an S-estimator [17] that deals a complex minimization problem to yields a robust estimate of the covariance matrix $\mathbf{C}$ . S-estimators can occasionally converge to an inaccurate local solution, so a good beginning point is needed. Despite, awaiting on MCD in the first stage confines Robust multivariate location and covariance matrix from investigating large data sets, particularly those of high dimension. It would appear that methods based on combinatorial search and alternative there of obtains an inherent inability to investigate large data sets.

### 2.2.3 Projection Pursuit

In adverse to distance-based procedures are projection pursuit methods [18] which can equivalently be applied to robust estimation as a main goal or carries towards explicit outlier detection. The fundamental purpose of projection pursuit methods is to find suitable projections of the data in which the outliers are readily credible and can thus be downweighted to turn out a robust estimator, which in turn can be used to detect the outliers. Because they do not consider the data to begin from a particular distribution but only search for useful projections, projection pursuit methods are not altered by non-normality and can be mainly applied in diverse data situations. The penalty for such independence comes in the form of increased computational stress, because it is not clear which projections should be tested; an exact method would require that all attainable directions be tested. The primal equivariant robust estimator having a high breakdown point in arbitrary dimension was the Stahel-Donoho estimator [19-20]. Computer approximation based on instructions from random subsamples was developed by Stahel (1981), but without any doubt a large amount of time is necessary to obtain acceptable results. Even though projection pursuit algorithms have the advantage of being appropriate in different data situations, their computational difficulties seem horrible.

## 2.3 The High Dimensional Situation

High dimensional data present several problems to classical statistical analysis. As earlier discussed, computation time increases more rapidly with $p$ than with $n$. For combinatorial and projection pursuit algorithms, this increase is of enough importance to put in question the practicability of such methods for high dimensional molecular descriptor data. In between the speedy distance-based methods, computation times of algorithms increase linearly with $n$ and cubically with $p$. This indicates that for very high dimensional molecular descriptor data, the computational stress of inverting the scatter matrix is nontrivial. This is particularly evident in iterative methods which require many iterations to converge, since the covariance matrix is inverted on each iteration. Thus, while the Mahalanobis distance is a very useful metric for finding correlated multivariate outliers, it is expensive to calculate. Every other methods of detection of outliers fare even worse, however, usually give up either computational time or detection accuracy. The MCD is a good case of this in that the precise solution is very accurate but infeasible to compute for all but small molecular descriptor data sets, whereas a faster solution can be obtained if random subsampling is used to produce an approximate solution. It will be investigated in results and discussion whether the subsampling version of MCD is competitive regarding both accuracy and computation time. Projection pursuit methods including the Stahel-Donoho estimator have computation times that increase very rapidly in higher dimensions, and are often at least an order of magnitude slower than distance based methods since their search for appropriate projections is a naturally time-consuming task. Thus even if the Mahalanobis distance may be computationally difficult due to the matrix inversion step, the robust version – RD, as defined in equation (1) – is an accurate metric for outlier detection and could well be more computationally fair than other ways.

This is appropriate to several biological applications where the data commonly have orders of weight more dimensions than observations. This is also the typical situation in chemometrics, which led to the development of Partial Least Squares (PLS) [21] with other methods. Because the covariance matrix is singular the robust Mahalanobis distance cannot be calculated. This is not as big a problem as initially occurs, after all, because the data can be transformed via singular value decomposition to an equivalent space of dimension $n − 1$ [2] and the analysis carried in the same way as $p < n$. However, this situation needs special attention and most outlier algorithms have to be modified to way this type of high-dimensional molecular descriptor data.

High dimensional molecular descriptor data have various interesting geometrical properties, discussed in [22]. One such property that is particularly appropriate to outlier detection, is that high dimensional data points lie near the surface of an enlarging sphere. For example , if $\|x\|$ is the norm of $x = (x_1, \ldots, x_n)^T$ drawn from a normal distribution with zero mean and identity covariance matrix, then, for ample $p$ we have

$$\frac{\|x\|}{\sqrt{p}} = \frac{\sqrt{\sum_{j=1}^{p}(x_j)^2}}{\sqrt{p}} \longrightarrow 1,$$

because the summation includes a $\chi_p^2$ distribution In this manner, if the outliers have even a slightly dissimilar covariance structure from the inliers(non-outliers), they will lie on a different sphere. This does not help low dimensional outlier detection, but if an algorithm is efficient of processing high dimensional molecular descriptor data, it should not be too hard to discover the different spheres of the outliers and inliers.

Principal components are a well known method of dimension reduction, that also advice an way to detecting high dimensional outliers. Recall that principal components are those directions that maximize the variance along each component, directed to the circumstance of orthogonality. Because outliers increase the variance along their respective directions, it appears instinctive that outliers will come into more visible in principal component space than the original data space; i.e. , Minimum some of the directions of maximum variance are hopefully to be those that enable the outliers to "tie-up" more. Exploring for outliers in principal component space should at least, till, not be any worse than searching for them in the original data space. If the data originally exist in a high-dimensional space, many of these dimensions likely do not provide significant additional information and are irrelevant. Principal components thus pick out a handful of highly informative components (relative to the total number of components), thereby performing a high degree of dimension reduction and making the data set much more computationally manageable without losing a lot of information.

For high dimension molecular descriptor data, a ample portion of the smaller principal components are actually noise [23]. Particularly if $p \ll n$ , the larger part of principal components will actually be noise and will not add to the total variance. By considering only those principal components that constitute some agreed level of the total variance, the number of components can be extensively reduced so that only those components that are truly meaningful are kept. Almost we found good results using a level of 99%. It can be argued this yields similar results to transforming the data via SVD to a dimension less than the minimum of $n$ and $p$. Thus, in place of imposing a level of improvement to the variance such as 99%, it would also be likely to choose the $n - 1$ (or fewer) components with the largest variance.

As marked in equation (7) in the OGK method [5] , after dividing by the MAD, the Euclidean distance in principal components space is for that reason similar to a robust Mahalanobis distance, because the off-diagonal elements of the scatter matrix are zero. Mahalanobis distance, because the off-diagonal elements of the scatter matrix are zero.

Hence, it is not essential to invert a p × p matrix when computing a measure of outlyingness for every point (i.e. the robust Mahalanobis distance), on the other hand slightly to divide (or "standardize") each principal component by its specific variance element. Because eigenvector decomposition has computational complexity $p^3$ to matrix inversion, doing the robust distance computations in principal component space is not more time-consuming than in common data space. If this transformation benefits the outliers become more visible and reduces the number of iterations required to detect them, the result will be a net savings in computational time.

It can be seen that the above approaches are based on simple basic properties of principal components; this is additional proof of how principal components continue to present appealing properties to both theoretical and applied statisticians.

## 2.4 Details of the  Proposed  PrCmpOut Method

The method we propose consists of two basic parts: First one aims to detect location outliers, and a next that aims to detect scatter outliers. Scatter outliers obtains a different scatter matrix than the rest of the data, while location outliers are described by a different location parameter. To begin, it is useful to robustly rescale or sphere each component using the coordinatewise median and the MAD, presenting to

$$x_{i,j}^* = \frac{x_{i,j} - med\big(x_{1,j}, \ldots \ldots, x_{nj}\big)}{MAD\big(x_{1,j}, \ldots \ldots, x_{n,j}\big)}, \quad j = 1, \ldots \ldots, p. \qquad (8)$$

Dimensions with a MAD of zero should be one or the other excluded , or alternatively another scale measure has to be used. Beginning with the rescaled data $x_{i,j}^*$ , we calculate a weighted covariance matrix, from which we compute the eigen values and  eigen vectors and hence a semi-robust principal components breakdown. We absorb only those eigen vectors or eigen values that provide to at least 99% of the total variance; call this new dimension  $p^*$. The remaining components are generally useless noise and d only deal to complicate any fundamental structure. For the case $p \gg n$, this also clarifies the  singularity problem because $p^* < n$. For the  $p^* \times p^*$ matrix $V$ of eigenvectors we thus get the matrix of principal components as

$$Z = X^* V, \qquad (9)$$

where $X^*$ is the matrix with the elements $x_{i,j}^*$ Rescale these principal components by the median and the MAD very much alike to equation (8),

$$z_{i,j}^* = \frac{z_{i,j} - med\big(z_{1,j}, \ldots \ldots, z_{nj}\big)}{MAD\big(z_{1,j}, \ldots \ldots, z_{n,j}\big)}, \quad j = 1, \ldots \ldots, p^*. \qquad (10)$$

Collect   $Z^*$   for the second stage of the algorithm. After the above pre-processing steps, the location outlier stage is initiated by computing the absolute value of a robust kurtosis measure for each component as stated in :

$$w_j = \left| \frac{1}{n} \sum_{i=1}^{n} \frac{\big(z_{ij}^* - med\big(z_{1j}^*, \ldots \ldots, z_{1n}^*\big)^4\big)}{MAD\big(z_{1j}^*, \ldots \ldots, z_{1n}^*\big)^4} - 3 \right|, \qquad j = 1, \ldots \ldots, p^* \qquad (11)$$

We make use of the absolute value because very much like to Multivariate outlier detection and robust covariance matrix estimation [24] , both small and large values of the kurtosis coefficient can be characteristics of outliers. This allows us to assign weights to each component according to how likely we think it is to disclose the outliers. We use respective weights $w_j / \sum_i w_i$  to present a well known scale $0 \leq w_j \leq 1$. If no outliers are there in a given component, we assume the principal components to be nearly normally distributed similar to the original data, producing a kurtosis near to zero. Because the being of outliers is likely to originate the kurtosis to become different than zero, we weight each of the $p^*$  dimensions proportional to the absolute value of its kurtosis coefficient. Assigning equal weights to all components (meanwhile the computation of robust Mahalanobis distances) reduce the strength of the discriminatory power because if outliers undoubtedly stick out in one component, the information in this component will be weakened unless it is given higher weight. Specifically in principal component space, outliers are more likely to be clearly visible in one specific component than slightly visible in several components, so it is important to assign this component higher weight. Because the components are uncorrelated, we compute a robust Mahalanobis distance employing the distance from the median (as scaled by the MAD), weighting each component according to the relative weights $w_j/\sum_i w_i$ with the kurtosis measure $w_j$ expressed in equation (11). The kurtosis measure expressed in equation (11) helps to guarantee that important information included in a particular component is not diluted by components which do not separate the outliers.

To complete the first stage of the algorithm, we want to determine how large the robust Mahalanobis distance should be to get an accurate classification between outliers and non-outliers. The kurtosis weighting strategy destroys any similarity to a $\chi^2_{p^*}$ distribution that might have been introduce , so it is not possible use a $\chi^2_{p^*}$ quantile as a separation barrier. Although , resembling to Maronna and Zamar (2002) and to equation (6), we got that transforming the robust distances $\{RD_i\}$ as stated in

$$d_i = RD_i \cdot \frac{\sqrt{\chi^2_{p^*,0.5}}}{med(RD_1, \ldots \ldots, RD_n)} \quad for \; i = 1, \ldots \ldots, n \qquad (12)$$

helped the experimental distances $\{d_i\}$ to have the same median as the assumed distances and thus bring the earlier somewhat near to $\chi^2_{p^*}$ , where $\chi^2_{p^*,0.5}$ is the $\chi^2_{p^*}$ 50th quantile. We make use of the adapted biweight function [25] to assign weights to every observation and use these weights as a measure of outlyingness. The adapted biweight fits into the general scheme of S-estimators defined by equation (4) and is similar to Tukey's biweight function other than that $\psi$ begins from 0 at some point $M$ from the origin. That is, observations closer than the scaled distance $M$ to location estimate accepts full weight of 1. The $\psi$ function of the adapted biweight is thus given by

$$\psi(d, c, M) = \begin{cases} d, & 0 \leq d < M \\ \left(1 - \left(\frac{d-M}{c}\right)^2\right)^2, & M \leq d \leq M + c \; , \\ 0, & d > M + c \end{cases} \qquad (13)$$

which corresponds to the weighting function

$$w(d; c, M) = \begin{cases} 1, & 0 \leq d < M \\ \left(1 - \left(\frac{d-M}{c}\right)^2\right)^2, & M \leq d \leq M + c \; , \\ 0, & d > M + c \end{cases} \qquad (14)$$

Precisely assigning known non-outliers full weights of one while assigning known outliers weights of zero results in increased capabilities of the estimators (supported the classifications are correct), and is also computationally faster. Between these peaks is a subset of points that receive weights similar to the usual biweight function. To support a high level of robustness, it is best to be moderate in assigning weights of one, since if any outliers enter the course of action with a weight of one (or close to it) that will make the other outliers harder to detect due to the masking effect. Remember that since the principal components have been scaled by the median and MAD these robust distances measure a weighted distance from the median (using transformed units of the MAD). We found good experimental values assigning a weight of one to the 1/3 of points acquiring the smallest robust distances. At the other end of the weighting pattern we assign zero weight to points with $d_i > c$, where

$$c = med(d_1, \ldots, d_n) + 2.5 \cdot MAD(d_1, \ldots, d_n) , \qquad (15)$$

corresponding nearly to classical outlier barriers. Similar to equation (14), the weights for each observation are calculated by the interpreted biweight function as stated in

$$w_{1i} = \begin{cases} 0, & d_i \geq c \\ \left(1 - \left(\dfrac{d_i - M}{c - M}\right)^2\right)^2, & M < d_i < c , \\ 1, & d_i \leq M \end{cases} \qquad (16)$$

where $i = 1, \ldots \ldots, n$ and $M$ is the $33\frac{1}{3}^{rd}$ of the quantile of the distances $\{d_1, \ldots, d_2\}$. Alternate weighing strategies were tested ; the benefit of the interpreted biweight is that it allows a subset of points (that we are quite definite are non-outliers) to be given full weight, while another subset of points that is likely to contain outliers can be given weights of zero, thereby excluding unacceptable influence by potential outliers, and a smooth weighting curve for in between points The weights. $\{w_{1i}\}$ from equation (16) are saved; we will use them again at the end of the algorithm. The second stage of our algorithm is similar to the first except that we don't use the kurtosis weighting layout. Principal components addresses on those directions that have large variance, so it is possibly not unexpected that we find good results searching for scatter outliers in the semi robust principal component space expressed at the beginning of this section. That is, we search for outliers in the space determined by $Z^*$ from equation (10). Earlier , computing the Euclidian norm for data in principal component space is similar to the Mahalanobis distance in the original data space, other than that it is faster to compute.

Because the distribution of these distances has not been modified like it was through the kurtosis weighting layout and assuming we start with normally distributed non-outliers, transforming the robust distances as before via equation (12) results in a distribution that is fairly close to $\chi^2_{p^*}$ . In establishing the interpreted biweight as in equation (16), then, acceptable results can be acquired by assigning $M^2$ equal to the $\chi^2_{p^*}$ 25$^{th}$ quantile and $c^2$ equal to the $\chi^2_{p^*}$ 99$^{th}$ quantile. This distribution is certainly not exactly equal to $\chi^2_{p^*}$, so there are chances when visual examination of these distances could head to a better boundary than this automated algorithm. Denote the weights calculated in this way, $w_{2i}$ , $i = 1, \ldots, n$.

Lastly we combine the weights from these two steps to compute eventual weights $w_i$, $i = 1, \ldots, n$, as stated in.

$$w_i = \frac{(w_{1i} + s)(w_{2i} + s)}{(1 + s)^2}, \qquad (17)$$

Where usually the scaling constant $s = 0.25$. The justification for presenting $s$ is that frequently too many non-outliers accept a weight of 0 in only one of the two steps; setting $s \neq 0$ helps to certify that the last weight $w_i = 0$ only if both steps assign a low weight. Outliers are then classified as points that have weight $w_i < 0.25$. These values means that if one of the two weights $w_{1i}$ or $w_{2i}$ equals one, the other must be less than 0.0625 for the point $x_i$ to be classified an outlier. Or, if $w_{1i} = w_{2i}$, then this common value must be less than 0.375 for $x_i$ to be classified as outlying.

We will hereafter call this algorithm as PrCmpOut. It is beneficial to recap the algorithm in brief.

**Stage 1:** Detection of location outliers

    **a)** Robustly sphere the data as stated in equation (8). Compute the sample covariance matrix of the transformed data $X^*$.

    **b)** Compute a principal component break down of the semi robust covariance matrix from Step a and retain only those $p^*$ eigenvectors whose eigenvalues provide to at least 99% of the total variance. Robustly sphere the transformed data as in equation (10).

    **c)** Compute the robust kurtosis weights for each component as in equation (11), and thus weighted norms for the sphered data from Step b. Because the data have been scaled by the MAD, these Euclidean norms in principal component space are similar to robust Mahalanobis distances. Convert these distances as stated in equation (12).

    **d)** Select weights $w_{1i}$ for every robust distance according the adapted biweight in equation (16), with $M$ equal to the $33\frac{1}{3}^{rd}$ quantile of the distances $\{d_1, \ldots, d_n\}$ and $c = med(d_1, \ldots, d_n) + 2.5 \cdot MAD(d_1, \ldots, d_n)$ .

**Stage 2:** Detection of scatter outliers

    **e)** Use the same semi-robust principal component break down computed in Step b and compute the (unweighted) Euclidean norms of the data in principal component space. Transform as stated in equation (12) to produce a set of distances for use in Step f.

    **f)** Select weights $w_{2i}$ for every robust distance as stated in the adapted biweight in equation (16) with $c^2$ equal to the $\chi_{p^*}^2$ 99th quantile and $M^2$ equal to the $\chi_{p^*}^2$ 25th quantile.

**Combining Stage 1 and Stage 2:** Use the weights from Steps d and f to select final weights for all observations as stated in equation (17).

**2.5 Preliminary Investigation of developed method**

As a preliminary step to developing a new outlier detection method, we briefly examined and compared existing methods to determine possible areas of improvement We selected nine algorithms that appeared to have good potential for finding outliers: rocke-Estimator [25], sfast-Estimator[26],M-Estimator[27],MVE-Estimator[28],NNC-Estimator [29] BACON[9],PCDist[10-11], sign1[12-13] and sign2[14]. Somewhat similar to our method, the Sign procedure is also based on a type of robust principal component analysis. It obtains robust estimates of location and spread based upon projecting the data onto a sphere. In this way, the effects of outlying observations are limited since they are placed on the boundary of the ellipsoid and the resulting mean and covariance matrix are robust. Standard principal components can thus be carried out on the sphered data without undue influence by any single point (or small subset of points). We considered attributes p = 10, 20 ,30,40 and critical values α= 0.05,0.1,0.15,0.2 , so for each parameter combination we carried out 14 simulations with n = 100 observations.

Frequently, results of this type are presented in a $2 \times 2$ table showing success and failure in identifying outliers and non-outliers, which we henceforth call inliers. In this paper, we present the percentage false negatives(FN) followed by the percentage false positives(FP) in the same cell; we henceforth refer to these respective values as the outlier error rate and inlier error rate(non outlier error rate). We propose these names because the outlier error rate specifies the percentage errors recorded within the group of true outliers, and similarly for the inlier error rate. True outliers among 100 observations are 10,16,18,22,23,25,27,29, 30,47,66,70,72,80,84 90,99 and 100. This is more compact than a series of $2 \times 2$ table and allows us to easily focus on bringing the error rates down to zero.

Table 1. Sample $2 \times 2$ table illustrating the notation used in this paper

|  | Predicted Outliers | Predicted Inliers |
|---|---|---|
| True Outliers | a | b |
| True Inliers | c | d |

$$Outlier\ Error\ rate\ (FN) = \frac{b}{a+b}\ ,Inlier\ Error\ rate\ (FP) = \frac{c}{c+d}$$

# 3. RESULTS AND DISCUSSION

Even if this algorithm was designed basically for computational performance at high dimension, we compare its performance against other outlier algorithms in low dimension, because a high dimensional comparison is not achievable. In the following we examine a variety of outlier configurations in molecular descriptor data. Examination of Table 2 unfolds that PrcCmpOut performs well at identifying outliers (low false positives), although it has a higher percentage of false negatives than most of the methods. PrCmpOut has the lowest percentage of false negatives, often by a tolerable margin, and is a competitive outlier detection method.

Table 2 Outlyingness measures: average percentage of outliers that were not identified and average percentage of regular observations that were declared outliers

| Critical Value (α) | p | Cutoff Value ($\chi^2_{p,1-\alpha}$) | Rocke | | M | | Sfast | | MVE | | NNC | | BACON | | PCDist | | Sign1 | | Sign2 | | PrCmpOut | |
|---|---|---|---|---|---|---|---|---|---|---|---|---|---|---|---|---|---|---|---|---|---|---|
| | | | FN (%) | FP (%) | FN (%) | FP (%) | FN (%) | FP (%) | FN (%) | FP (%) | FN (%) | FP (%) | FN (%) | FP (%) | FN (%) | FP (%) | FN (%) | FP (%) | FN (%) | FP (%) |
| 0.05 | 10 | 4.278672 | 22 | 29 | 44 | 34 | 67 | 15 | 39 | 30 | 39 | 23 | 39 | 21 | 72 | 13 | 33 | 17 | 44 | 13 | 50 | 8 |
| | 20 | 5.604501 | 44 | 31 | 44 | 31 | 51 | 10 | 33 | 28 | 28 | 20 | 17 | 24 | 78 | 13 | 33 | 17 | 72 | 3 | 44 | 7 |
| | 30 | 6.616115 | 56 | 24 | 44 | 26 | 44 | 12 | 44 | 28 | 11 | 13 | 17 | 23 | 89 | 13 | 28 | 20 | 89 | 0 | 39 | 6 |
| | 40 | 7.46716 | 67 | 25 | 67 | 26 | 39 | 14 | 44 | 25 | 22 | 8 | 44 | 14 | 89 | 8 | 22 | 26 | 72 | 0 | 33 | 0 |
| 0.1 | 10 | 3.998397 | 22 | 28 | 33 | 28 | 67 | 20 | 33 | 31 | 28 | 24 | 28 | 25 | 72 | 15 | 33 | 17 | 44 | 13 | 50 | 18 |
| | 20 | 5.330289 | 33 | 26 | 33 | 26 | 50 | 17 | 33 | 28 | 5 | 13 | 5 | 29 | 78 | 13 | 28 | 17 | 72 | 4 | 44 | 9 |
| | 30 | 6.344763 | 28 | 24 | 28 | 24 | 39 | 14 | 28 | 26 | 5 | 21 | 5 | 28 | 89 | 13 | 22 | 30 | 89 | 0 | 39 | 6 |
| | 40 | 7.197573 | 61 | 24 | 67 | 20 | 39 | 19 | 39 | 28 | 22 | 18 | 44 | 23 | 89 | 8 | 17 | 34 | 83 | 0 | 28 | 0 |
| 0.15 | 10 | 3.81234 | 28 | 30 | 28 | 30 | 67 | 20 | 28 | 31 | 28 | 25 | 22 | 26 | 72 | 15 | 44 | 19 | 39 | 15 | 50 | 15 |
| | 20 | 5.14758 | 39 | 25 | 33 | 26 | 28 | 14 | 11 | 29 | 5 | 17 | 5 | 32 | 72 | 13 | 17 | 21 | 67 | 6 | 39 | 10 |
| | 30 | 6.163623 | 28 | 24 | 28 | 25 | 33 | 17 | 17 | 25 | 5 | 23 | 5 | 34 | 83 | 13 | 17 | 36 | 83 | 0 | 39 | 6 |
| | 40 | 7.017396 | 50 | 21 | 50 | 25 | 33 | 19 | 33 | 28 | 11 | 20 | 33 | 21 | 89 | 9 | 17 | 37 | 83 | 0 | 22 | 0 |
| 0.2 | 10 | 3.666328 | 33 | 37 | 33 | 37 | 39 | 21 | 28 | 34 | 22 | 28 | 17 | 25 | 72 | 15 | 28 | 21 | 39 | 17 | 39 | 14 |
| | 20 | 5.003749 | 33 | 28 | 33 | 25 | 22 | 17 | 22 | 32 | 5 | 20 | 5 | 37 | 72 | 13 | 17 | 25 | 61 | 7 | 39 | 9 |
| | 30 | 6.020813 | 50 | 30 | 50 | 30 | 33 | 18 | 22 | 31 | 5 | 25 | 5 | 37 | 83 | 13 | 17 | 37 | 83 | 0 | 22 | 4 |
| | 40 | 6.875212 | 50 | 23 | 44 | 23 | 33 | 24 | 33 | 34 | 13 | 26 | 33 | 23 | 89 | 9 | 15 | 40 | 83 | 0 | 11 | 0 |

We present simulation results in which the dimension was increased from $p = 10$ to $p = 40$, based on the mean of 16 simulations at each level. In contrast to the previous simulation experiment in dimension $p = 10$, in this case we were not able to examine the performance of the other algorithms since they were not computationally feasible for these dimensions. The number of observations was held constant at $n = 100$, as was the number of outliers at 18. With increasing dimension PrCmpOut can identify almost all outliers. None of the known methods experience much success in identifying outliers for small dimensions because geometrically, the outliers are not very different from the non-outliers. However, as dimension increases, it can be seen how the outliers separate from the non-outliers and become easier to detect. At p = 20 dimensions, barely more than half of the outliers can be detected, at p = 30 dimensions almost 90% are detected, and at $p = 40$ dimensions more than 95% of the outliers are detected.

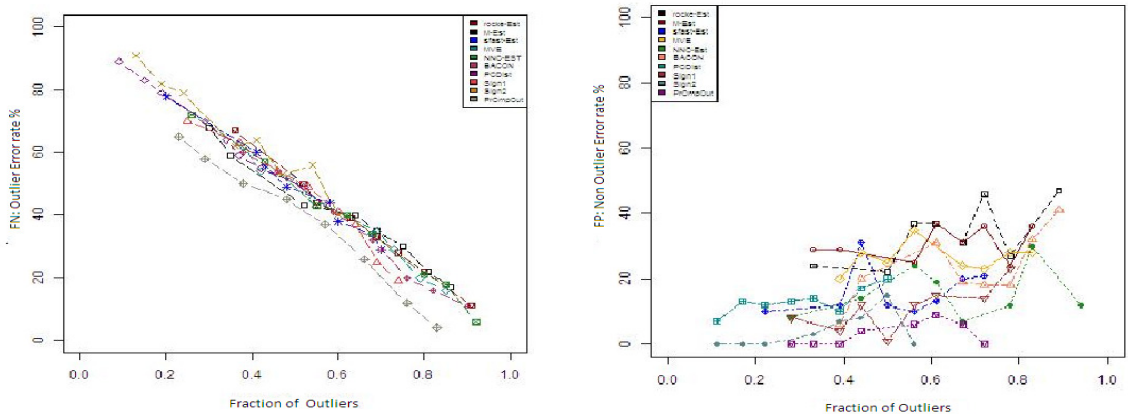

Figure 1. Average outlier error rate(left) and Average non-outlier error rate(right) for varying fraction of Outliers

The results for the PrCmpOut and rest of the estimators are presented in Figure 1. For an outlier fraction of 10% all estimators except PCDist perform excellent in terms of outlier error rate (FN) and detect outliers independently of the percentage of outliers as seen from the left panel of Figure 1. The average percentage of non-outliers that were declared outliers (FP) differ and PrCmpOut performs best, followed closely by Sign2 ,PCDist (below 10). With more than 20% comes next for the rest of the estimators. MVE declares somewhat more than 30% of regular observations as outliers.Sign1 performs worst with the average non-outlier error rate increasing with increase of the fraction of non outliers as an outliers.In terms of the outlier error rate PrCmpOut performs best (11%) followed by NNC, SIGN1, BACON and MVE Estimators error rate between 13% and 34% and PCDist (89%) is again last.

Exploratory data analysis is often used to get some understanding of the data at hand, with one important aspect being the possible occurrence of outliers. Sometimes the detection of these outliers is the goal of the data analysis, more often however they must be identified and dealt with in order to assure the validity of inferential methods. Most outlier detection methods are not very robust when the multinormality assumption is violated, and in particular when outliers are present. Robust distance plots are commonly used to identify outliers. To demonstrate this idea, we first compute the robust distances based on the sample mean vector and the sample covariance matrix. Points which have distances larger than $\chi^2_{p,1-\alpha}$ are usually viewed as potential outliers, and so we will label such points accordingly. Figure 2 shows the distance plots of all the methods.

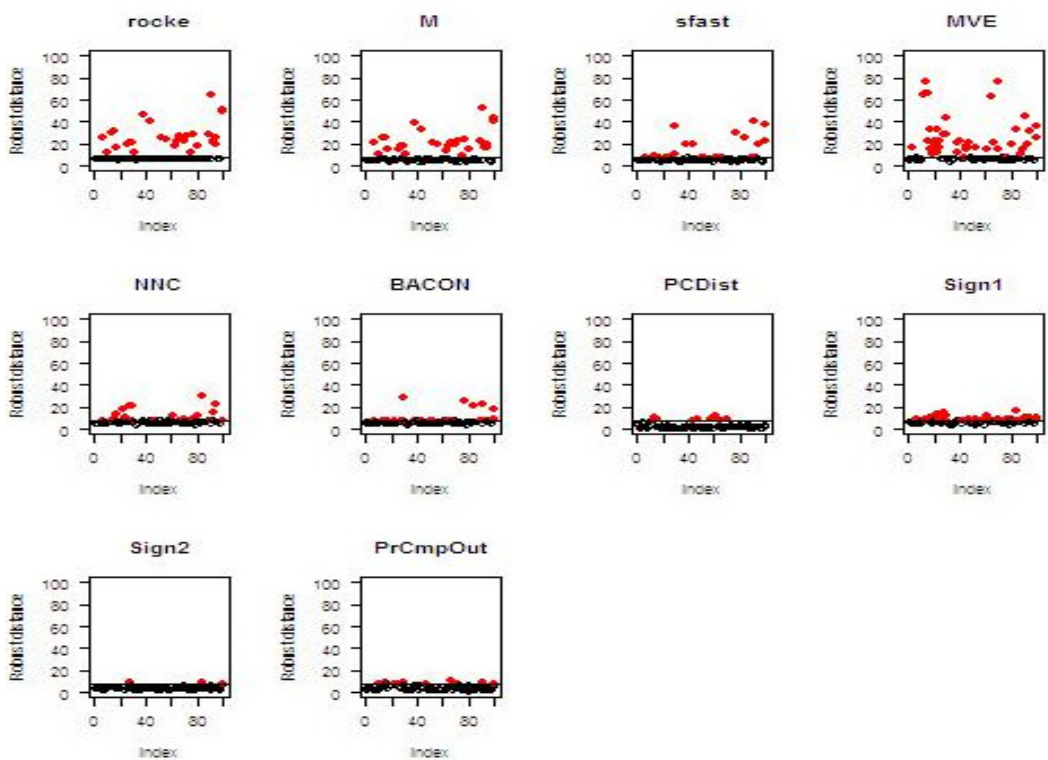

Figure 2. Robust distance plots for the Oxazolines and Oxazoles molecular descriptor data set. The red points are according to the robust distances outliers.

It is important to consider the computational performance of the different outlier detection algorithms in case of large Oxazolines and Oxazoles molecular descriptor dataset with n = 100. To evaluate and compare the computational times a simulation experiment was carried out. The experiment was performed on a Intel core i3 with 6Gb RAM running Windows 7 Professional. All computations were performed in R i386 2.15.3. The rocke, M,sfast, MVE, NNC, BACON, PCDist, Sign1, Sign2 and Proposed algorithm PrCmpOut algorithms from were used. Fastest is PrCmpOut, followed closely by Sign1,Sign2,PCDist ,NNC and BACON. Slowest is sfast-Est followed by rocke-Est, M-Est and MVE. These computation times are presented graphically in Figure 3 using a seconds scale

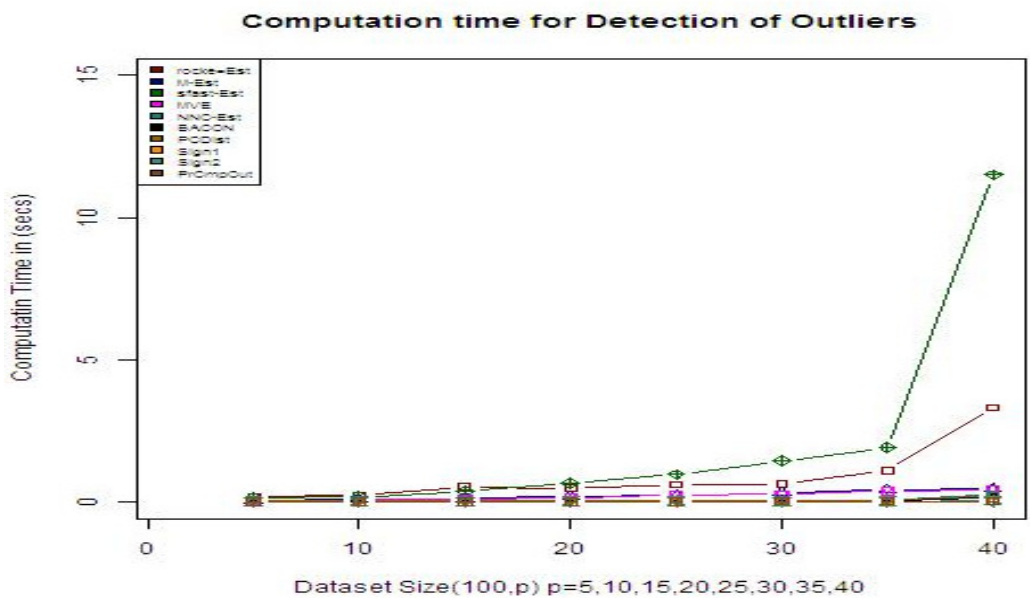

Figure 3. Comparing the computation time of outlier detection algorithms in the presence of outliers

A fast multivariate outlier detection method is particularly useful in the field of bioinformatics where hundreds or even thousands of molecular descriptors need to be analyzed. Here we will focus only on outlier detection among the molecular descriptors, clearly a high dimensional data set. First, columns with MAD equal to zero were removed, with the remaining columns investigated for outliers. Figure 4 shows the results from PrCmpOut; The intermediate graphs provide more insight into the workflow of the method: the kurtosis weights of Step 3 of the PrCmpOut algorithm are shown in the upper left panel, together with the weight boundaries described in Step 4, leading to the weights in the upper right panel. Similarly, the distances from Step 5 and the weights from Step 6 are respectively shown in the left and right panel of the second row. The lower left panel shows the combined weights (Phase 1 and 2 combined) together with the outlier boundary 0.25, which results in 0/1 weights (lower right panel). The observations are clearly visible as multivariate outliers in the intermediate steps of the algorithm

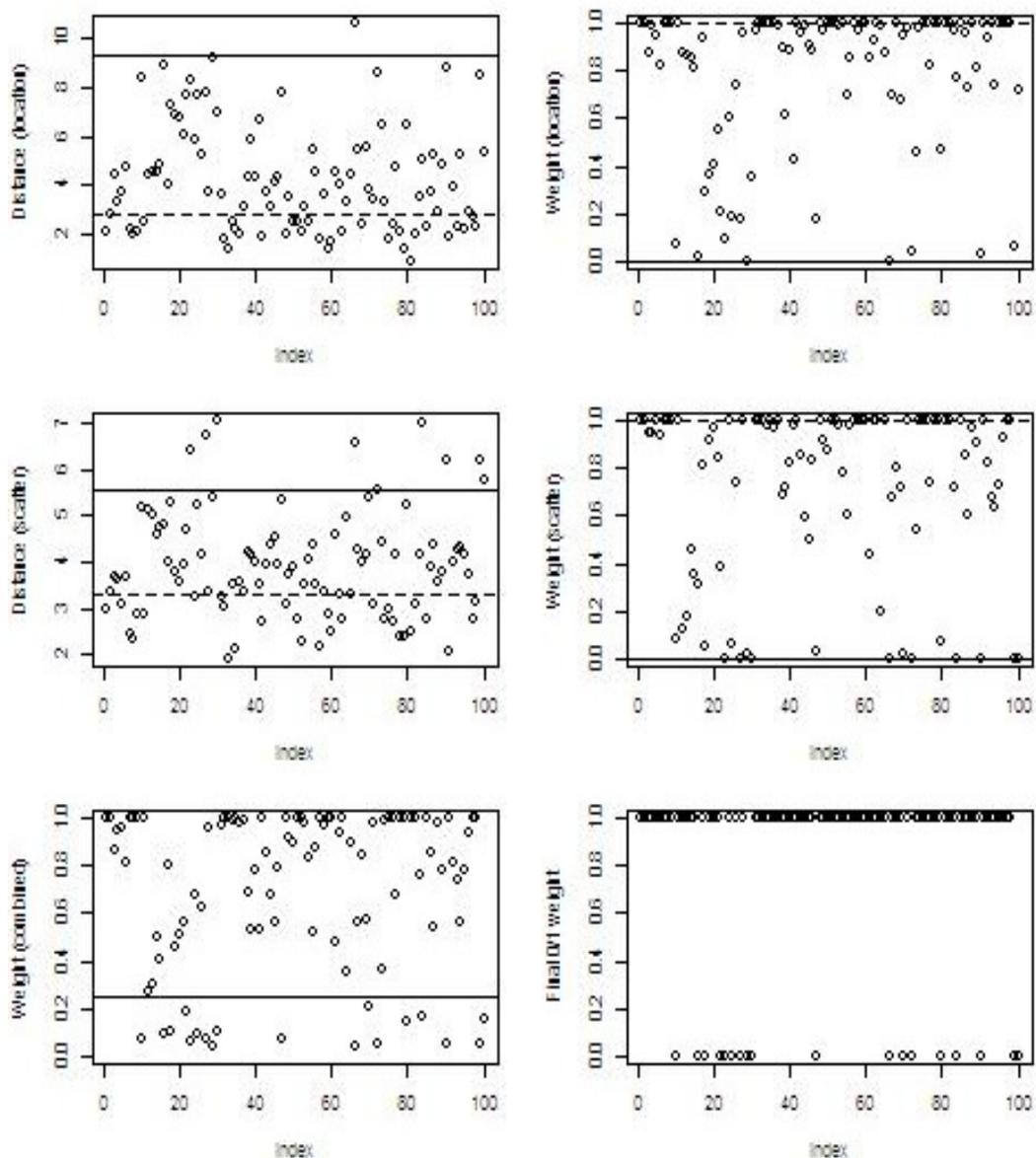

Figure 4. The panels show the intermediate steps of the PrCmpOut algorithm (distances and weights) for analyzing Oxazolines and Oxazoles molecular descriptor dataset.

We compare the performance of PrCmpOut on this data set with the Sign2 method; the other algorithms are not feasible due to the high dimensionality. Figure 5 shows the distances (left) and weights (right) as calculated by the Sign2 method.

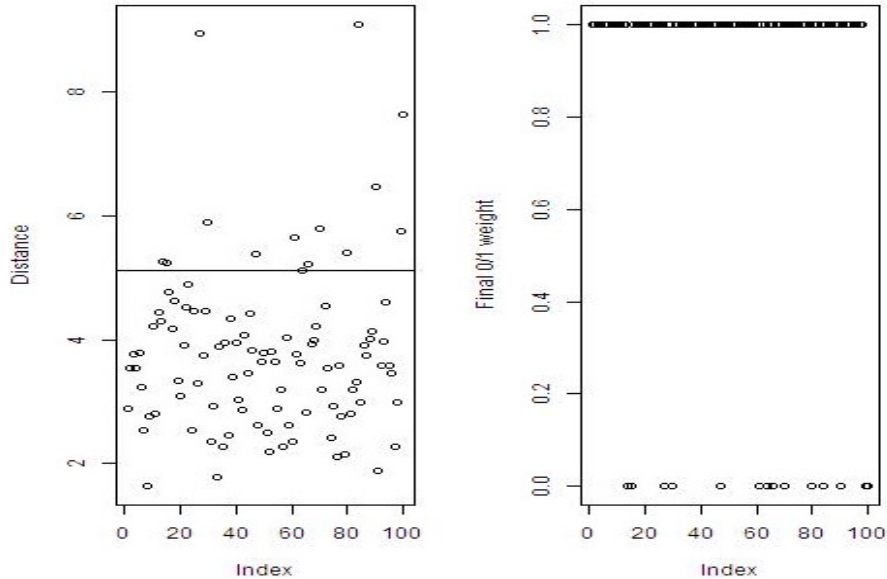

Figure  9. Distances and weights for analyzing the   Oxazolines and Oxazoles molecular descriptor dataset with the Sign method.

A possible explanation for the difficulty experienced by the Sign2 method is a masking effect for PCA. It is evident that PrCmpOut has better performance than the Sign2 method, which is also evident from the results in Table 2. We infer that PrCmpOut is a competitive outlier detection algorithm regarding detection accuracy as well as computation time.

## 4. CONCLUSIONS

PrCmpOut is a method for detecting outliers in multivariate data that utilizes inherent properties of principal components decomposition. It demonstrates very good performance for high dimensional data and through the use of a robust kurtosis measure. In this paper we tested several approaches for identifying outliers in Oxazolines and Oxazoles molecular descriptor dataset. Two aspects seem to be of major importance: the computation time and the accuracy of the outlier detection method. For the latter we used the fraction of false negatives(FN) - outliers that were not identified - and the fraction of false positives(FP) - non-outliers that were declared as outliers. It is very fast to compute and can easily handle high dimensions. Thus, it can be extended to fields such as bioinformatics and data mining where computational feasibility of statistical routines has usually been a limiting factor. At lower dimensions, it still produces competitive results when compared to well-known outlier detection methods.

### ACKNOLDGEMENTS

We gratefully thank to the Department of Computer Science Mangalore University, Mangalore India for technical support of this research.

# Authors

Doreswamy received B.Sc degree in Computer Science and M.Sc Degree in Computer Science from University of Mysore in 1993 and 1995 respectively. Ph.D degree in Computer Science from Mangalore University in  the year 2007. After completion of his Post-Graduation Degree, he subsequently joined and served as Lecturer in Computer Science at St. Joseph's College, Bangalore  from 1996-1999.Then he has elevated to the position  Reader in Computer Science at  Mangalore  University in year 2003. He was the Chairman of the Department of  Post-Graduate Studies and  research in computer science from 2003-2005 and from 2009-2008 and served at varies 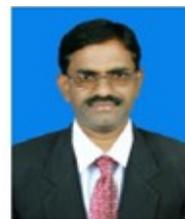

capacities in  Mangalore University at present he is the Chairman  of Board of Studies and Associate Professor  in Computer Science of Mangalore University. His areas of  Research interests include Data Mining and Knowledge Discovery, Artificial Intelligence and  Expert Systems, Bioinformatics, Molecular modelling and simulation, Computational Intelligence, Nanotechnology, Image Processing and Pattern recognition. He has been granted a Major Research  project entitled "Scientific Knowledge Discovery Systems (SKDS) for Advanced Engineering Materials Design Applications" from the funding  agency University Grant Commission, New Delhi, India. He has been published about 30 contributed peer reviewed Papers at  national/International  Journal  and  Conferences. He received SHIKSHA RATTAN PURASKAR for his outstanding achievements in   the  year 2009 and   RASTRIYA VIDYA SARASWATHI AWARD for outstanding  achievement in chosen field of  activity  in the  year 2010.

Chanabasayya M. Vastrad received B.E. degree and M.Tech. degree in the year 2001 and 2006  respectively. Currently working towards his Ph.D Degree in Computer Science and Technology  under the guidance of  Dr. Doreswamy in the Department  of  Post-Graduate Studies  and  Research  in  Computer Science , Mangalore University. 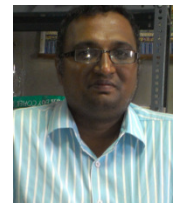